\documentclass[noeprint,noshowpacs,nopreprintnumbers,twocolumn,prb,showpacs,superscriptaddress,amsmath,amssymb,citeautoscript,aps,10pt]{revtex4-1}
\usepackage{graphicx}
\usepackage{dcolumn}
\usepackage{color}
\usepackage{bm}
\usepackage[hidelinks]{hyperref}
\hypersetup{
    colorlinks,
    citecolor=blue,
    filecolor=blue,
    linkcolor=blue,
    urlcolor=blue
}
\usepackage{lipsum}
\usepackage{amsmath}

\allowdisplaybreaks

\begin{document}

\title{Identities and Many-Body Approaches in Bose-Einstein Condensates}

\author{Shohei Watabe}
\email{shoheiwatabe@rs.tus.ac.jp}
\affiliation{Faculty of Science Division I, Department of Physics, Tokyo University of Science}

\date{\today}

\begin{abstract}
This paper discusses exact relations in Bose--Einstein condensates (BECs), starting from basic properties of an ideal Bose gas. 
In particular, focused on are the Hugenholtz--Pines relation, Nepomnyashchii--Nepomnyashchii identity, and identities for the density response function. 
After introducing these exact relations, a few approaches of many-body approximations are discussed, which satisfy the exact relations in BECs. This paper will serve as a bridge between theories on exact relations and those on  approximations in BECs. 
\end{abstract}

\maketitle

\section{Introduction}

Bose--Einstein condensation is a dramatic phenomenon, where quantum mechanics emerge in a macroscopic scale. 
Bose--Einstein condensates (BECs) provide rich variety of topics in condensed matter physics, such as superfluidity, phase transition, symmetry breaking and so on~\cite{AndersonBook,LeggettBook,PitaevskiiBook}. 
These topics are strongly related to other condensed matter physics such as superconductors as well as magnetisms~\cite{AndersonBook,LeggettBook}. 
The BEC is thus a cornucopia of basic concepts of condensed matter physics.

During a long history of the study of BECs, beautiful identities have been shown~\cite{Hugenholtz1959,Gavoret1964,Nepomnyashchii1975,Nepomnyashchii1978,GriffinBook}. 
These are related to the gapless excitation~\cite{Hugenholtz1959}, the infrared divergence of the longitudinal susceptibility~\cite{Nepomnyashchii1975,Nepomnyashchii1978}, which is strongly related to the magnetism~\cite{Vaks1967,Patashinskii1973}, and the correspondence between sound speed of single-particle excitation and that of the collective density excitation in a BEC~\cite{Gavoret1964}.
On the other hand, from a view point of a more practical approximation side, not only the difficult problem still remains for the dilemma of the conserving-gapless approximation ~\cite{Hohenberg1965}, but the treatment of the infrared divergence is also difficult~\cite{Shi1998}. 
We furthermore still often encounter gaps between known exact properties and approximation frameworks that do not satisfy them. 

This paper may serve as a bridge that connects these gaps, by starting to  introduce basic properties of an ideal Bose gas. 
We then introduce exact relations for an interacting condensed Bose gas, such as the Hugenholtz--Pines relation~\cite{Hugenholtz1959}, Nepomnyashchii--Nepomnyashchii identity~\cite{Nepomnyashchii1975,Nepomnyashchii1978}, as well as the exact relations with respect to the density response function~\cite{Gavoret1964}. 
Finally, we discuss many-body approaches satisfying these identities~\cite{Watabe2013,Watabe2014,Watabe2018}. 
Development of approximation frameworks for BEC is not completed, in the sense that there is not a practical approximation satisfying all the exact relations in BECs. 
We hope that this paper helps to guide readers towards theories on BECs, and is useful for developing them.

\section{Quantum Statistical Mechanics of Bosons}

An ideal classical gas is well described by the canonical ensemble. At very low temperature, however, the treatment of a classical gas is no longer valid. 
Its temperature scale is described by using one of the typical length scales of a nonzero temperature system --- the thermal de Broglie length $\lambda _T \equiv [ 2\pi \hbar^2 / (m k_{\rm B} T) ]^{1/2}$, where $T$ is the temperature, and $m$ is an atomic mass of a gas~\cite{LeggettBook}. 
When this thermal de Broglie length has a same or longer length scale of the mean interatomic distance $d$, i.e., $\lambda_T \gtrsim d$, the treatment of the classical gas becomes failure, and the quantum statistical mechanics is needed. 

Indeed, the partition function $Z$ of a three-dimensional ideal classical gas is given by $Z = \zeta^N/N !$ with the Boltzmann counting factor $N!$, where 
\begin{align} 
	\zeta \equiv \frac{1}{h^3} \int d {\bf p} \int d {\bf r} \exp \left ( - \beta \epsilon_{\bf p} \right ) 
	= \frac{V}{\lambda_T^3}.  
\end{align}
Here, $\beta$ is the inverse temperature $\beta = 1/(k_{\rm B} T)$, $\epsilon_{\bf p} \equiv {\bf p}^2 / (2m)$ is a kinetic energy of a particle in a uniform system, $V$ is the volume of the system. 
The entropy $S = - \partial F / \partial T$ with the Helmholtz free energy $F = - \beta^{-1} \ln Z$ is given by 
\begin{align} 
	S = k_{\rm B} N \left [ \frac{5}{2} + \ln \left ( \frac{d}{\lambda_T} \right )^3 \right ], 
\end{align}
where the mean interatomic distance $d \equiv (V/N)^{1/3}$~\cite{GreinerBook}. 
Now that we may find that an unphysical negative entropy is shown in the case $\lambda_T \gtrsim d$, we should not consider the classical gas treatment in this region. Instead, we consider the quantum statistical mechanics for identical particles at very low temperature $k_{\rm B} T \lesssim 2\pi \hbar^2 /(md^2)$. 

This statistics may be found by the symmetry of the multiparticle wavefunction. 
The two-particle wave function for two identical particles at a time $t$ is given by $\Psi ({\bf r}_1, {\bf r}_2; t) $. Since the two particles are identical, we have a relation $| \Psi ({\bf r}_1, {\bf r}_2; t) | = | \Psi ({\bf r}_2, {\bf r}_1; t) |$. 
This gives a symmetry relation~\cite{LeggettBook}
\begin{align}
	 \Psi ({\bf r}_1, {\bf r}_2; t) = \pm  \Psi ({\bf r}_2, {\bf r}_1; t) , 
\end{align}
where the upper sign is for boson with integer spin, and the lower sign is for fermion with half-integer spin. 
In contrast to the fermion case, where $\Psi ({\bf r}, {\bf r} ; t) = 0$ for ${\bf r} \equiv {\bf r}_1 = {\bf r}_2$, which provides the anti-bunching effect, the bosons may give $\Psi ({\bf r}, {\bf r} ; t) \neq 0$, which provides the bunching effect for bosons. The fermion obeys the Pauli exclusions principle, where a single quantum state cannot be occupied by two or more fermions. On the other hand, a single quantum state can be occupied by any numbers of bosons. 

In the grand canonical formulation, the Bose distribution function --- the mean occupation number of a quantum state $i$ for boson --- is given by~\cite{PitaevskiiBook} 
\begin{align}
	 f (\epsilon_i) = \frac{1}{\exp [\beta (\epsilon_i - \mu)] -1 }, 
\end{align}
where $\epsilon_i$ is an energy of a single-particle state $i$, and $\mu$ is the chemical potential that satisfies $\epsilon_0 > \mu$. Here, $\epsilon_0$ is the lowest single-particle energy. 
Using a density of states for a three-dimensional uniform system 
\begin{align}
	\rho (\epsilon ) \equiv \sum\limits_{\bf p} \delta (\epsilon - \epsilon_{\bf p}) 
	= \frac{V m^{3/2}}{\sqrt{2} \pi^2 \hbar^3} \sqrt{\epsilon}, 
\end{align}
the number of particles without assuming the BEC is given by 
\begin{align}
	N = \sum\limits_{\bf p} f (\epsilon_{\bf p}) =  \int_0^\infty d \epsilon \rho (\epsilon ) f (\epsilon) 
	= \frac{V}{\lambda_T^3} {\mathcal G}_{3/2} (z), 
\end{align}
where $z$ is the fugacity $z \equiv \exp(\beta \mu)$, and 
\begin{align}
	{\mathcal G}_\alpha (z) \equiv \frac{1}{\Gamma (\alpha) } \int_0^\infty d x \frac{x^{\alpha-1} }{\exp (x) z^{-1} - 1} 
	= \sum\limits_{n=1}^\infty \frac{z^n}{n^\alpha}
\end{align}
with the Gamma function $\Gamma (\alpha)$. 

By introducing the phase space density $\rho_{\rm sd} \equiv N \lambda_T^3 /V$, we have the relation 
$\rho_{\rm sd} = {\mathcal G}_{3/2} (z)$. 
Since the relation $\epsilon_{{\bf p} = 0} = 0 > \mu$ holds, which leads $0 <  z < 1$, we have the following relation in the limit $z \to 1$, given by 
\begin{align}
	\rho_{\rm sd}^{} = \frac{N}{V} \lambda_T^3 = {\mathcal G}_{3/2} (z \to 1) = \zeta(3/2). 
\end{align}
With lowering temperature or increasing density, however, we should have a high phase space density with $\rho_{\rm sd}^{} =N \lambda_T^3 / V> \zeta(3/2)$. 
In this case, the relation $N = \lambda_T^{-3} V {\mathcal G}_{3/2} (z) $  is no longer valid, 
and we need to reconsider the relation 
\begin{align}
	N = f (\epsilon_{{\bf p} = 0}) + \sum_{{\bf p} \neq 0} f (\epsilon_{\bf p})  \equiv N_0 + N'. 
\end{align}
The second term with $\mu \to 0$, i.e., $z \to 1$, gives 
\begin{align}
	N' = & \sum_{{\bf p} \neq 0} f (\epsilon_{\bf p}) 
	= \lim\limits_{\delta \to 0} \int_\delta^\infty \rho (\epsilon) f (\epsilon) = \frac{V}{\lambda_T^3} \zeta(3/2),  
\end{align}
which is intensive, where $N'$ increases proportionally to $V$. 
The first term $f (\epsilon_{{\bf p} = 0})$ with $\mu \neq 0$ may be negligible compared with the intensive second term $N'$ in the thermodynamic limit $V\to \infty$. 
However, if the chemical potential approaches to zero in the limit $V \to \infty$ with the relation 
\begin{align}
	\mu = - \frac{k_{\rm B} T}{n_0 V}, 
\end{align}
the number of particles occupying the lowest energy state $N_0 = f (\epsilon_{{\bf p} = 0})$ may become intensive in the thermodynamic limit $V, N\to \infty$ with $N/V = {\rm const.}$, which provides $f (\epsilon_{{\bf p} = 0}) \simeq k_{\rm B} T/|\mu| = n_0 V$. 
The coefficient $n_0$ is here found to be the condensate density given by $n_0 = N_0 / V$. 
This is the Bose--Einstein condensation, where the lowest energy state is macroscopically occupied, 
and its critical temperature for an ideal Bose gas $T_{\rm c}^0$ is given by 
\begin{align}
	k_{\rm B} T_{\rm c}^0 \equiv \frac{2\pi \hbar^2}{m} \left [ \frac{N}{V} \frac{1}{\zeta(3/2)} \right ]^{3/2}.
\end{align}

\section{Bogoliubov Prescription}

We now consider the non-zero temperature Green's function formalism. 
We start with a non-interacting Bose gas, where the Hamiltonian is given by 
\begin{align}
	\hat H_0 = & \int d {\bf r} \left [ 
	\frac{\hbar^2}{2m} \nabla \hat \Psi^\dag ({\bf r}) \nabla \hat \Psi ({\bf r}) 
	- \mu \hat \Psi^\dag ({\bf r}) \hat \Psi ({\bf r}) 
	\right ]. 
\end{align}
Here, $\hat \Psi ({\bf r})$ and $\hat \Psi^\dag ({\bf r})$ are an annihilation and creation operator of a bosonic atom at a spatial position ${\bf r}$, satisfying a commutation relation $[ \Psi ({\bf r}), \Psi^\dag ({\bf r}')] = \delta ({\bf r} - {\bf r}')$. 
The imaginary time Green's function is a very powerful tool for studying physics at nonzero-temperatures, which is defined as 
\begin{align}
	G_0 ({\bf r}, \tau; {\bf r}', \tau') \equiv 
	& - \langle T_\tau \hat \Psi ({\bf r}, \tau) \hat \Psi^\dag ({\bf r}', \tau') \rangle 
	\\ 
	\equiv & 
	- \frac{{\rm Tr} [ \exp (- \beta \hat H_0 ) \hat \Psi ({\bf r}, \tau) \hat \Psi^\dag ({\bf r}', \tau') ] }{
	{\rm Tr} [ \exp (- \beta \hat H_0  )]
	}, 
\end{align}
where $\tau$ is the imaginary time. 
The real space representation of the annihilation operator $\hat \Psi ({\bf r}, \tau)$ and its momentum space $\hat a_{\bf p} (\tau)$ are related through 
\begin{align}
	\hat \Psi ({\bf r}, \tau) = \cfrac{1}{\sqrt{V}} \sum\limits_{\bf p} \exp \left ( \frac{i {\bf p} \cdot {\bf r} }{ \hbar } \right ) \hat a_{\bf p} (\tau). 
\end{align}
The non-interacting Green's function in the momentum and Matsubara frequency spaces is given by~\cite{StoofBook}
\begin{align} 
G_0({\bf p}, i\omega_n )= & - \int_0^{\hbar \beta} d \tau \exp (i \omega_n \tau) 
\langle T_\tau \hat a_{\bf p} (\tau ) \hat a_{\bf p}^\dag (0) \rangle 
\\ 
= & \cfrac{\hbar}{i \hbar \omega_n - \epsilon_{\bf p} + \mu}, 
\end{align}
where $T_\tau$ is the time ordering operator, where the operator at smaller imaginary time is ordered at the right, and $\omega_n = 2 \pi n /(\hbar \beta)$ with $n \in \mathbb{Z}$ is the bosonic Matsubara frequency. 
This non-zero temperature formalism can be mapped to the absolute zero temperature case with the analytic continuation $i \omega_n \to \omega + i \eta$ with an infinitesimally small number $\eta$. 
The pole of the Green's function $G_0^{-1} (p) = 0$ provides the dispersion relation of the excitation, 
where the ideal gas case provides $\hbar \omega = \epsilon_{\bf p} - \mu$. 
By taking the sum of the Matsubara frequency and using the counter integration, we obtain the Bose distribution function for an ideal Bose gas, given by~\cite{StoofBook}
\begin{align}
	f (\epsilon_{\bf p} )
	= & 
	- \frac{1}{\hbar \beta} \lim\limits_{\eta\to 0} \sum\limits_n 
	\exp{(i \omega_n \eta)} G_0 ({\bf p}, i\omega_n ) . 
	\\ 
	= & \cfrac{1}{\exp [ \beta (\epsilon_{\bf p} - \mu)] -1} . 
\end{align}

One of the treatments for the Bose--Einstein condensation is the Bogoliubov prescription, where the field operator of the zero-momentum state is replaced by a $c$-number, given by 
\begin{align}
	\hat \Psi ({\bf r}) = 
	 & \frac{\hat a_{{\bf p} = 0}}{\sqrt{V}} 
	+ 
	\frac{1}{\sqrt{V}} \sum\limits_{{\bf p} ( \neq 0)} \exp (i {\bf p} \cdot {\bf r}/\hbar) \hat a_{\bf p}
	\\ 
	\to & \Phi_0 + 
	\frac{1}{\sqrt{V}} \sum\limits_{{\bf p} ( \neq 0)} \exp (i {\bf p} \cdot {\bf r}/\hbar) \hat a_{\bf p}. 
\end{align}
Here, $\Phi_0$ is the condensate wavefunction, which is related to the condensate density $n_0$ and the phase $\varphi_0$ through $\Phi_0 = \sqrt{n_0} \exp (i \varphi_0)$. 

Since the field operators of the zero-momentum part satisfy~\cite{AbrikosovBook,FetterBook}
\begin{align} 
\left [ \frac{ \hat a_{{\bf p} = 0}} {\sqrt{V}} , 
\frac{ \hat a_{{\bf p} = 0}^\dag } {\sqrt{V}}
\right ] = \frac{1}{V}, 
\end{align}
the error of the replacement of the operator with the $c$-number may be given by ${\mathcal O} (1/V)$, which is expected to be negligibly small in the thermodynamic limit $V \to \infty$.
In the BEC phase, the density matrix 
$\rho_{(1)} ({\bf r}_1, {\bf r}_2) = N_0/V - G_0 ({\bf r}_1 , \tau; {\bf r}_2, \tau + \eta)$ has the off-diagonal long-range order, which can be given by~\cite{PitaevskiiBook}
\begin{align}
	\lim\limits_{|{\bf r}_1 - {\bf r}_2| \to \infty }
	\rho_{(1)} ({\bf r}_1, {\bf r}_2)
	= \frac{N_0}{V}. 
\end{align} 

\section{identities of interacting condensed Bose gas}

We now consider an interacting Bose gas, where the Hamiltonian $\hat H$ is given by 
\begin{align}
\hat H = \hat H_0 + \frac{U}{2} \int d {\bf r} 
	\hat \Psi^\dag ({\bf r}) \hat \Psi^\dag ({\bf r}) \hat \Psi ({\bf r}) \hat \Psi ({\bf r}). 
\end{align}
The interaction strength $U$ is related to an $s$-wave scattering length $a$ through a relation
\begin{align}
	\frac{4 \pi a}{m} = & \frac{U}{ 1 + U \sum\limits_{\bf p}^{p_{\rm c}} 1/(2 \epsilon_{\bf p}) }, 
\end{align}
where $p_{\rm c}$ is a cutoff momentum.
In the BEC phase, we apply the Bogoliubov prescription to the field operator. 
In the Green's function formalism, interaction effect is included through the self-energy ${\bf \Sigma} (p)$, and the Green's function in the BEC phase is given by the Dyson--Beliaev equation, where~\cite{AbrikosovBook,FetterBook}
\begin{align}
	{\bf G} (p) = {\bf G}_0 (p) + {\bf G}_0 (p) {\bf \Sigma} (p) {\bf G} (p), 
	\label{eq27}
\end{align}
with 
\begin{align}
	{\bf G} (p)  = 
	\begin{pmatrix}
		G_{11} (p) & G_{12} (p) \\ G_{21} (p) & G_{22} (p)
	\end{pmatrix}, 
	{\bf \Sigma} (p)  = 
	\begin{pmatrix}
		\Sigma_{11} (p) & \Sigma_{12} (p) \\ \Sigma_{21} (p) & \Sigma_{22} (p)
	\end{pmatrix}, 
\end{align}
and ${\bf G}_0 (p) = {\rm diag} [G_0 (p), G_0 (-p)]$. 

We here note the lowest contribution of ${\bf G} (p)$. 
By using the Dyson--Beliaev equation, we have the infinite series of the Green's function ${\bf G} (p) = {\bf G}_0 (p) + {\bf G}_0 (p) {\bf \Sigma} (p) {\bf G}_0 (p)  + {\bf G}_0 (p) {\bf \Sigma} (p) {\bf G}_0 (p) {\bf \Sigma} (p) {\bf G}_0 (p) + \cdots$, and 
the lowest contribution of the normal Green's function $G_{11} (p)$ is found to be $G_{11} (p) \simeq G_0 (p) + \cdots$, which means that the first order contribution is the non-interacting Green's function, which gives $G_{11} (p) = {\mathcal O} (U^0)$. 
On the other hand, the lowest contribution of the anomalous Green's function is given by $G_{12} (p) \simeq G_0 (p) \Sigma_{12} (p) G_0 (-p) + \cdots$. 
In the lowest contributions of the self-energy $\Sigma_{12} (p) = U \Phi_0^2$, we can find that the anomalous Green's function has $G_{12} (p) = {\mathcal O} (U^1)$. The order of the contribution of the interaction is different between the normal and anomalous Green's functions. 
 
The Dyson--Beliaev equation gives the full-Green's function ${\bf G}(p) = \hbar / [{\bf G}_0^{-1} (p) - \hbar {\bf \Sigma} (p)]$, 
which provides~\cite{AbrikosovBook,FetterBook} 
\begin{align}
	G_{11} (p) = & 
	\frac{\hbar}{D(p)} [ \hbar \omega + \epsilon_{\bf p} - \mu + \hbar \Sigma_{11} (-p) ], 
	\label{eq29}
	\\ 
	G_{12} (p) = & 
	- \frac{\hbar^2 \Sigma_{12} (p)}{D(p)}, 
	\label{eq30}
\end{align}
where 
\begin{align}
	D (p) \equiv  & 
	[\hbar \omega - \hbar A(p)]^2 
	- 
	\left [ \epsilon_{\bf p} - \mu + \hbar S(p) + \hbar \Sigma_{12} (p) \right ]
	\nonumber
	\\ 
	& \times  
	\left [ \epsilon_{\bf p} - \mu + \hbar S(p) - \hbar \Sigma_{12} (p) \right ] , 
\end{align}
with 
\begin{align}
	A (p) \equiv & [\Sigma_{11} (p) - \Sigma_{11} (-p)]/2, 
	\\ 
	S (p) \equiv & [\Sigma_{11} (p) + \Sigma_{11} (-p)]/2. 
\end{align}
Here, we used $\Sigma_{11} (p) = \Sigma_{22} (-p)$ and $\Sigma_{12} (p) = \Sigma_{21} (-p)$. 
The same relation also holds for the Green's function $G_{ij} (p)$. 

The Hugenholtz--Pines relation~\cite{Hohenberg1965}
\begin{align}
\mu = \hbar \Sigma_{11} (0) - \hbar \Sigma_{12} (0)
\label{Hugenholtz-Pines-eq}
\end{align}
ensures that an excitation of the single-particle excitation in a BEC is gapless. 
Indeed, (\ref{Hugenholtz-Pines-eq}) gives the denominator of (\ref{eq29}) and (\ref{eq30}) being zero, i.e., $D (p = 0) = 0$. 
This relation can be proven by various ways. 
One of the key ideas of the proof comes from the fact that the energy of the system is gauge invariant~\cite{Hugenholtz1959}.  
A key idea of another proof comes from the linear response of the bosonic field operator with respect to an infinitesimally small symmetry breaking external field~\cite{Hohenberg1965}. 
Using the Hugenholtz--Pines relation, the single-particle Green's function in the low-energy limit is given by~\cite{Gavoret1964} 
\begin{align} 
G (p) \simeq & \frac{n_{0} m c^{2}}{n} \frac{ \hbar }{ \hbar^2 \omega_{}^{2} - c^{2} {\bf p}^{2} } \begin{pmatrix} 1 & -1 \\ -1 & 1 \end{pmatrix} , 
\label{eq70} 
\end{align} 
where $c$ is the sound speed of the single particle excitation in the low-energy limit, and $n \equiv N/V$ is the total number density of particles. 

The sound speed $c$ of the single particle excitation in (\ref{eq70}) is equal to the thermodynamic sound speed, given by~\cite{Gavoret1964,GriffinBook}
\begin{align}
\frac{n}{mc^2} = \left ( \frac{dn}{d\mu}\right )_T. 
\label{eq37}
\end{align}
It indicates that the sound speed of the single-particle excitation is equal to that of the density mode given by the density-density correlation function,~\cite{Gavoret1964,GriffinBook} 
\begin{align}
\chi (p) \simeq \frac{n}{m} \frac{{\bf p}^2}{\hbar^2 \omega^2 - c^2 {\bf p}^2}. 
\label{eq38}
\end{align} 
Indeed, the pole of this correlation function is originally comes from that of the single-particle Green's function~\cite{Gavoret1964,GriffinBook}. 
These structures are specific to the BEC system, and important points are (i) the single-particle Green's function is involved to the density correlation function because of the existence of the condensate, and (ii) the single-particle Green's function has a pole giving the phonon dispersion relation, whose sound speed is the same as the thermodynamic sound speed. 

The Nepomnyashchii--Nepomnyashchii identity~\cite{Nepomnyashchii1975,Nepomnyashchii1978}
\begin{align}
\Sigma_{12} (0) = 0
\end{align}
is another identity, which is caused by the non-vanishing weak infrared divergence of the higher order correlation functions.
This identity provides the weak infrared divergence of the longitudinal susceptibility~\cite{Nepomnyashchii1983,GriffinBook}, which is also related to the fact that the main contribution of the excitation in the low-energy limit is the phase fluctuation.

In this paper, the details of the proofs for these exact properties are not shown. However, some useful ideas for them will be summarized. 

For the Hugenholtz--Pines theorem, it is important that the system energy is gauge invariant, and the normal and anomalous self-energies in the limit $p \to 0$ is generated from the system energy $E$ per volume by eliminating the condensate wave-function~\cite{Hugenholtz1959,Nepomnyashchii1978}. 
In the BEC phase, the system energy is given by the sum of the connected diagrams constructed by $G_0(p)$ as well as $\Phi_0^{(*)}$. 
Since the system energy is gauge invariant, it does not depend on the choice of the phase of the field operator as well as the condensate wave function. 
Indeed, the normal Green's function $G_0(p) = - \langle T_\tau \hat a_{\bf p} (\tau ) \hat a _{\bf p}^\dag (0)\rangle$, which has the same numbers of $\hat a_{\bf p}$ and $\hat a_{\bf p}^\dag$, is unchanged by the gauge transformation $(\hat a_{\bf p}, \hat a_{\bf p}^\dag) \to ( \exp (i \varphi) \hat a_{\bf p}, \exp (-i \varphi) \hat a_{\bf p}^\dag ) $. 
In order to keep the system energy gauge invariant, a certain contribution to the system energy should have the same number of $\Phi_0$ and $\Phi_0^*$, which provides the system energy is also unchanged by the gauge transformation $(\Phi_0, \Phi_0^* ) \to ( \exp (i \varphi) \Phi_0, \exp (-i \varphi) \Phi_0^* ) $. 
If a condensate wave function $\Phi_0$ or $\Phi_0^*$ is eliminated from the energy functional, a single vertex point of $p = 0$ is generated. 
This idea can be easily extended to generating multi-points vertex functions. 
Suppose that a certain contribution to  the energy functional $E^{(s)}$ has $\Phi_0$ and $\Phi_0^*$, whose numbers are $s$, respectively. 
The normal self-energy is generated by eliminating a single $\Phi_0$ from $s$ possibilities of $\Phi_0$ in $E^{(s)}$, and by eliminating a single $\Phi_0^*$ from $s$ possibilities of $\Phi_0^*$ in $E^{(s)}$. 
On the other hand, the anomalous self-energy is generated by eliminating a single $\Phi_0^*$ from $s$ possibilities of $\Phi_0^*$ in $E^{(s)}$, and then eliminating a single $\Phi_0^*$ from the remaining $s-1$ possibilities of $\Phi_0^*$. 
As a result, we have relations 
\begin{align}
\hbar \Sigma_{11}^{(s)} (0) =  \frac{s}{\Phi_0} \frac{s}{\Phi_0^*} E^{(s)} 
, 
\quad 
\hbar \Sigma_{12}^{(s)} (0) =  \frac{s}{\Phi_0^*} \frac{s-1}{\Phi_0^*} E^{(s)}, 
\label{eq39}
\end{align}
which leads 
\begin{align}
\hbar \Sigma_{11}^{(s)} (0) - \exp (-i 2 \varphi_0) \hbar \Sigma_{12}^{(s)} (0) = 
& \frac{s}{n_0} E^{(s)} . 
\end{align}
The right hand side, on the other hand, is given by~\cite{Hugenholtz1959}. 
\begin{align}
\frac{s}{n_0} E^{(s)} = \frac{\partial E^{(s)}}{\partial n_0} \equiv \mu^{(s)}.
\end{align}
By employing the Bogoliubov prescription, we may have a term $- \mu_0 |\Phi_0|^2 = - \mu_0 n_0$ satisfying $\mu = \mu_0$ in the original energy $\bar E (T, \mu, \mu_0)$. 
However, the contribution $- \mu_0 n_0$ is not generated from the energy functional constructed by the Green's function, which instead gives $E (T, \mu, n_0) = \bar E (T, \mu, \mu_0) + \mu_0 n_0$. 
This is a kind of the Legendre transformation, and we find $dE = - SdT - p dV - n' d\mu + \mu_0 dn_0$, which gives $\mu = \mu_0 = \partial E / \partial n_0$. 
After collecting all the possible contributions with respect to $s$, we have the Hugenholtz--Pines relation 
\begin{align}
\hbar \Sigma_{11} (0) - \exp (-i 2 \varphi_0) \hbar \Sigma_{12} (0) = 
\mu. 
\label{HPeq2}
\end{align}
Since the phase of the condensate wave function is often taken as to be $\varphi_0 = 0$, 
Eq. (\ref{HPeq2}) can be reduced to Eq. (\ref{Hugenholtz-Pines-eq}), 
where (\ref{Hugenholtz-Pines-eq}) is found to be a specific representation for $\Phi_0 = \sqrt{n_0}$. 

From (\ref{eq39}), we can generate the relations~\cite{Nepomnyashchii1978} 
\begin{align}
\hbar \Sigma_{11}^{} (0) = \frac{\partial}{\partial n_0} \left ( n_0 \frac{\partial E}{\partial n_0} \right ), 
\hbar \Sigma_{12}^{} (0) = e^{i 2 \varphi_0} n_0 \frac{\partial^2 E}{\partial n_0^2}. 
\label{eq39prime}
\end{align}
Since the phase $\varphi_0$ may be taken arbitrary, we take the average of self-energies for the phase of the condensate wave-function, i.e., $\langle \Sigma_{ij} (0) \rangle = (2\pi)^{-1} \int_0^{2\pi} d \varphi_0 \Sigma_{ij} (0) $. 
It is clearly seen that the anomalous self-energy is gauge dependent, and we can find $\langle \Sigma_{12} (0) \rangle = 0$~\cite{Watabe2014}. 
On the other hand, the normal self-energy is gauge independent, and $\langle \Sigma_{11} (0) \rangle \neq 0$. 
It is also the case for the chemical potential. 
This is one of the simplest ways for understanding the Nepomnyashchii--Nepomnyashchii identity $\Sigma_{12} (0) = 0$. 

This identity is also strongly related to the weak infrared divergence of the longitudinal susceptibility, where we here introduce the longitudinal and transverse susceptibilities~\cite{Weichman1988,Giorgini1992}, 
\begin{align}
\chi_\nu (p) = \int_0^{\hbar \beta} d \tau 
e^{i \omega_n \tau} \langle T_\tau 
\hat a_{\nu, {\bf p}} (\tau) 
\hat a_{\nu, -{\bf p}} (0) 
\rangle. 
\end{align}
Here, 
$\hat a_{\perp,{\bf p}} = (\hat a_{\bf p} - \hat a_{-{\bf p}}^\dag)/(2i)$, 
and 
$\hat a_{\parallel,{\bf p}} = (\hat a_{\bf p} + \hat a_{-{\bf p}}^\dag)/2$ 
are transverse and longitudinal operators, respectively. 
The transverse fluctuation is consistent with the phase fluctuation. 
Although the longitudinal operator may be often referred to as the density fluctuation or the amplitude fluctuation, those are not exactly the same and the careful treatment of those operators are needed, because the longitudinal operator $\hat a_{\parallel} = (\hat a + \hat a^\dag )/2$ has the gauge dependence. 
The amplitude mode, known as the Higgs mode, may be detected by the scalar susceptibility, not the longitudinal susceptibility~\cite{Podolsky2011}. 
In the low-energy regime, we have relations 
\begin{align}
	\chi_{\perp} (0, {\bf p}) \simeq \frac{n_0 m}{n|{\bf p}|^2}, 
	\quad 
	\chi_{\parallel} (0, {\bf p}) \simeq \frac{1}{4 \Sigma_{12} (0, {\bf p})}. 
	\label{eq45}
\end{align}
If the Nepomnyashchii--Nepomnyashchii identity $\Sigma_{12} (0) = 0$ holds, the infrared divergence of the longitudinal susceptibility emerges.

The Popov's hydrodynamic theory is known as an absolute-zero temperature approximation that satisfies both the Hugenholtz--Pines relation and the Nepomnyashchii--Nepomnyashchii identity~\cite{PopovBook,Popov1972,Popov1979}. 
In this theory, the bosonic field operator $\hat \Psi (x) \equiv \sqrt{n_0 + \hat \pi(x)} e^{i \hat \varphi (x) }$ is expanded by hydrodynamic operators $\hat \pi$ and $\hat \varphi$~\cite{PopovBook,Popov1972,Popov1979,Dupuis2011}, 
and the Green's function ${\bf G}$ is approximately given by the sum of correlation functions $G_{\hat \pi \hat \pi}, G_{\hat \pi \hat \varphi}, G_{\hat \varphi \hat \pi}, G_{\hat \varphi \hat \varphi}$ as well as the convolution of $G_{\hat \varphi \hat \varphi}$. 
By inversely solving the Dyson--Beliaev equation (\ref{eq27}) for ${\bf \Sigma}$ with the given ${\bf G}$ in the Popov's hydrodynamic theory, we can find the self-energy satisfies both the Hugenholtz--Pines relation and the Nepomnyashchii--Nepomnyashchii identity~\cite{Popov1979}. 
In this formalism, the leading order contribution of the longitudinal susceptibility is the convolution of $G_{\hat \varphi \hat \varphi} \propto 1/(\hbar^2 \omega^2 - c^2 |{\bf p}|^2)$~\cite{Popov1979}. 
This convolution in the low-energy regime shows the weak infrared divergence, given by~\cite{GriffinBook}
\begin{align} 
\chi_{\parallel} (p) \propto & 
\sum\limits_q G_{\hat \varphi \hat \varphi} (p+q) G_{\hat \varphi \hat \varphi} (q) 
\\ 
\propto & 
\begin{cases}
\ln (c^2 |{\bf p}|^2 - \hbar^2\omega^2) & (T = 0)
\\ 
1/ |{\bf p}| & (T \neq 0). 
\end{cases}
\label{eq47}
\end{align} 
As a result, the anomalous self-energy vanishes as $\Sigma_{12} \simeq 1/(4 \chi_{\parallel}) \propto |{\bf p}|$ in the limit ${\bf p} \to 0$ at $T \neq 0$. 

Appearance of the convolution of the transverse-field (or phase) correlation in the longitudinal susceptibility is originated from the fact that the excitation in the low-energy limit is exhausted by the transverse field (or the phase of BEC), and the amplitude of the order-parameter may not change in this limit~\cite{Patashinskii1973}. 
Let $\Phi_0 = \sqrt{n_0}$ be the original field, and in the low-energy excitation with the fluctuation, the field may be approximately given by 
\begin{align} 
\Phi_0 ' \simeq & (\sqrt{n_0}+ \delta \Phi_\parallel)\exp(i \delta \Phi_\perp) 
\\ 
\simeq & \Phi_0 + \delta \Phi_\parallel + i \sqrt{n_0} \delta \Phi_\perp. 
\end{align} 
If the amplitude of the order parameter does not change in the low-energy limit, i.e., $|\Phi_0 '|^2 = |\Phi_0 |^2$, 
we have a condition 
\begin{align} 
\delta \Phi_\parallel \simeq - \frac{\sqrt{n_0}}{2} \delta \Phi_\perp^2 . 
\end{align} 
As a result, the leading order of the longitudinal correlation function $\langle \delta \Phi_\parallel \delta \Phi_\parallel \rangle$ is given by the transverse fluctuations, i.e, $\langle \delta \Phi_\perp^2 \delta \Phi_\perp^2 \rangle$, which provides the convolution of the transverse correlation function $\langle \delta \Phi_\perp \delta \Phi_\perp \rangle$. 
This story is consistent with the infrared divergence of the longitudinal susceptibility and that of the Popov's hydrodynamic theory. 

The original derivation of the Nepomnyashchii--Nepomnyashchii identity~\cite{Nepomnyashchii1975,Nepomnyashchii1978} is different from those approximation theories. The self-energies --- two-point vertices --- can be generated from the three-point vertices, where two of three vertex points are connected to two single-particle Green's functions, which are also connected to an interaction line as well as a single condensate line ~\cite{Nepomnyashchii1978}. 
Although almost all infrared divergences are canceled out each other, some contributions remains, which is the same contribution as that in (\ref{eq47})~\cite{Nepomnyashchii1978,Watabe2014}. 
On the other hand, as discussed in the topic of the Hugenholtz--Pines identity, the vertex functions in the zero-energy limit is generated from the system energy by eliminating the condensate wave-functions. 
As a result, three-point vertices in the low-energy limit can be related to the two-point vertices (i.e., self-energies), and one can obtain an equation with respect to the anomalous self-energy. 
By solving this equation, one can find the anomalous self-energy vanishes in the zero-momentum and zero-energy limits, which is caused by the inverse of the weak-infrared divergence shown in (\ref{eq47})~\cite{Nepomnyashchii1978,Watabe2014}. 

As related to the fact that the self-energy in the zero-energy limit can be generated from the energy functional, the self-energy with the small but nonzero momentum and that with the small but nonzero energy is also generated by the similar way~\cite{Gavoret1964}. 
The self-energy contribution in the low-energy regime is thus related to the thermodynamic quantities. Indeed, we have a relation~\cite{Gavoret1964} 
\begin{align}
	\frac{1}{\hbar} 
	\left [ 
	\frac{\partial^2 \Sigma_{11} (0) }{\partial \omega^2}
	- 
	\frac{\partial^2 \Sigma_{12} (0) }{\partial \omega^2}
	\right ] 
	= \frac{1}{n_0}
	\frac{\partial^2 E}{\partial \mu^2}
	= - \frac{n}{n_0 m c^2}, 
	\label{eq51}
\end{align}
where we used relations $n' = - \partial E' / \partial \mu$, and $\partial n' / \partial \mu = d n / d\mu= n /(m c^2)$. 
Here, $c$ is the thermodynamic sound speed. 
The thermodynamic functional is constructed by the non-interacting single-particle Green's function $G_0$, the condensate wavefunction $\Phi_0$ and $\Phi_0^*$, as well as the interaction parameter $U$. 
If one increase the energy of $G_0$ with the infinitesimally small value, i.e., $G_0 (p) = \hbar / [ \hbar (\omega + \delta \omega ) - \epsilon_{\bf p} + \mu ]$, we may regard it as the infinitesimally small increment of the chemical potential, which provides the relation $\hbar^{-1} \partial G_0 / \partial \omega = \partial G_0 / \partial \mu$~\cite{Gavoret1964}. 
The second derivative of the self-energy with respect to the frequency can be thus related to the second derivative of the energy functional with respect to the chemical potential. 
Using (\ref{Hugenholtz-Pines-eq}) and (\ref{eq51}), one can show that the sound speed of the single-particle excitation in the low-energy limit is given by the thermodynamic sound speed, as discussed in (\ref{eq37}) and (\ref{eq38}). 

The density response function in the BEC phase is constructed by two contributions, one-particle reducible (1PR) part, and one-particle irreducible (1PI) part~\cite{Gavoret1964}. 
The 1PR part is specific to the BEC phase, given by 
\begin{align}
	\chi^{\rm 1PR} (p) = {\bf \Upsilon}^\dag (p) {\bf G} (p) {\bf \Upsilon} (p), 
\end{align}
where ${\bf \Upsilon}$ and ${\bf \Upsilon}^\dag$ are the density vertices that include the single-particle Green's function into the density response function and vanish above the critical temperature. 
This vertex is known to show the zero-frequency density vertex identity~\cite{Nepomnyashchii1978} 
\begin{align}
	{\bf \Upsilon} (0) = 0, 
\end{align}
and the 1PR part in the low energy regime is given by~\cite{Gavoret1964}
\begin{align}
	\chi^{\rm 1PR} (p) = \frac{n}{mc^2} \frac{(\hbar\omega)^2}{(\hbar \omega)^2 - c^2 {\bf p}^2}. 
\end{align}
This implies that $\lim_{{\bf p} \to 0} \chi^{{\rm 1PR}} (0, {\bf p} ) = 0$~\cite{Gavoret1964}. 
On the other hand, the 1PI part in the low-energy limit exactly shows~\cite{Gavoret1964} 
\begin{align}
	\chi^{\rm 1PI} (0) = - \frac{n}{mc^2}. 
\end{align}
Since the density response function is the sum of the 1PR and 1PI parts, i.e., $\chi = \chi^{\rm 1PR} + \chi^{\rm 1PI}$, 
the compressibility zero-frequency sum-rule $\chi(0) = - n/(mc^2)$ is exhausted by the 1PI part.


\section{Many-body treatment}

This section focuses on many-body approaches at non-zero temperatures, such as the random-phase approximation, and the many-body $T$-matrix theory~\cite{Watabe2013,Watabe2014,Watabe2018}. 
For simplicity, we take the convention $V = \hbar = k_{\rm B} = 1$ in this section.  
It may be convenient to construct building blocks for many-body contributions from the single-particle Green's function in the Hartree--Fock--Bogoliubov--Popov approximation (Shohno model)~\cite{GriffinBook}, given by 
\begin{align}
	g (p) = 
	\begin{pmatrix}
		g_{11} & g_{12} \\ g_{21} & g_{22}
	\end{pmatrix}
	= 
	\frac{1}{i \omega_n \sigma_3 - \xi_{\bf p} - U n_0 \sigma_1} , 
\end{align}
where $\xi_{\bf p} = \epsilon_{\bf p} + U n_0$. Here, $\sigma_{1,2,3}$ are the Pauli matrices. 
This gives the Bogoliubov excitation --- the gapless phonon excitation in the low-energy limit ---, which captures properties of the exact single-particle Green's function. 
One of the building blocks is the correlation function, given by~\cite{Watabe2013,Watabe2014,Watabe2018} 
\begin{align}
	\Pi (p) = - T \sum\limits_q g(p+q) \otimes g(-q), 
\end{align}
where $\otimes$ is the Kronecker product. 
Since the normal and anomalous Green's function has the opposite sign in the low-energy regime, 
i.e., $g_{ij} (p) \simeq (-1)^{i+j} mc_0^2 /(\omega^2 - c_0^2 {\bf p}^2)$, where the Bogoliubov phonon sound speed is $c_0 = \sqrt{U n_0/m}$, 
the infrared divergence of this correlation function $\Pi$ has a simple structure, which can be extracted by the following matrix~\cite{Watabe2013,Watabe2014,Watabe2018}
\begin{align}
	\Pi^{\rm IR} (p) = 
	\Pi_{14} (p) \hat C , 
\end{align}
where 
\begin{align}
\hat C = 
\begin{pmatrix}
	1 & -1 & - 1 & 1 
	\\ 
	- 1 & 1 & 1 & - 1 
	\\ 
	- 1 & 1 & 1 & - 1
	\\
	1 & -1 & - 1 & 1
\end{pmatrix}, 
\end{align}
with $\Pi_{14} (p) = - T \sum\limits_q g_{12}(p+q) g_{12} (-q)$. 
The function $\Pi_{14}$ shows the same structure of the infrared divergence in (\ref{eq47}), where $c$ is replaced by $c_0$. 
The correlation function $\Pi^{\rm R} \equiv \Pi - \Pi^{\rm IR}$ converges to a finite value in the low-energy limit. 

Using this correlation function $\Pi$, 
the four point vertex $\Gamma$ and the regular part of the density response function $\chi_{\rm R}$ are respectively given by~\cite{Watabe2013,Watabe2014,Watabe2018} 
\begin{align}
	\Gamma (q) = & \frac{U}{1-U\Pi(q)}, 
	\\ 
	\chi_{\rm R} (q) = & \frac{1}{2} \langle f_0 | [\Pi (q) + \Pi (q) \Gamma(q) \Pi (q) ] | f_0 \rangle, 
	\label{eq61}
\end{align}
where $\langle f_0 | = (0,1,1,0)$ and $| f_0 \rangle = (0, 1,1, 0)^{\rm T}$. 
This regular part does not show the infrared divergence and converges to~\cite{Watabe2013,Watabe2014,Watabe2018} 
\begin{align}
	\chi_{\rm R} (0) = & -\frac{1}{U} \frac{ 1 - U \Pi' (0) }{ 2 - U \Pi ' (0) }, 
\end{align}
with 
\begin{align}
\Pi' (0) \equiv \sum\limits_{{\bf p}} \frac{\epsilon_{\bf p}^2}{E_{\bf p}^2} 
	\left ( 
		\frac{\partial n_{\bf p}}{\partial E_{\bf p}}
		- 
		\frac{1 + 2 n_{\bf p}}{2 E_{\bf p}}
	\right ). 
\end{align}

The random-phase approximation (RPA) includes the density fluctuation into the effective interaction, where the effective interaction $U_{\rm eff}$ is given by~\cite{Watabe2013,Watabe2018} 
\begin{align}
	U_{\rm eff} (q) = \frac{U}{1-U \chi_{\rm R} (q)} . 
\end{align}
Using this interaction, we consider the self-energies as 
\begin{align}
	\Sigma_{11} (p) = & (n_0 + n_0' ) U_{\rm eff} (0) + n_0 U_{\rm eff} (p) 
	\nonumber 
	\\ & 
	- T \sum\limits_q U_{\rm eff} (q) g_{11} (p-q) ,  
	\label{eq65}
	\\ 
	\Sigma_{12} (p) = & n_0 U_{\rm eff} (p), 
	\label{eq66}
\end{align}
where $n_0' \equiv - T \sum_p g_{11} (p) e^{i\omega_n \delta}$. 
Since $\chi_{\rm R} (0)$ does not show the infrared divergence, we find $U_{\rm eff} (0) \neq 0$. 
This random-phase approximation provides $\Sigma_{12} (0) = n_0 U_{\rm eff} (0) \neq 0$, which does not satisfy the Nepomnyashchii--Nepomnyashchii identity. 

The variant of this RPA, which satisfies the Nepomnyashchii--Nepomnyashchii identity, is the simplified RPA (s-RPA), where we include the lowest order of the contribution~\cite{Watabe2013,Watabe2018} 
\begin{align}
	\chi_{\rm R}^{0} (q) = & \frac{1}{2} \langle f_0 | \Pi (q) | f_0 \rangle. 
\end{align}
The effective interaction in this s-RPA is given by 
\begin{align}
	U_{\rm eff}^0 (p) = \frac{U}{1-U\chi_{\rm R}^0 (p) }. 
\end{align}
Self-energies are also given by replacing $U_{\rm eff}$ with $U_{\rm eff}^0$ in (\ref{eq65}) and (\ref{eq66}). 
Since $\chi_{\rm R}^0$ shows the infrared divergence, the effective interaction $U_{\rm eff}^0$ converges to zeros $U_{\rm eff}^0 (0) = 0$. As a result, the anomalous self-energy satisfies the Nepomnyashchii--Nepomnyashchii identity $\Sigma_{12} (0) = n_0 U_{\rm eff}^0 (0) = 0$. 

The many-body $T$-matrix (MBT) approximation is given by~\cite{Watabe2013}  
\begin{align}
	\Sigma_{11} (p) = & 2n_0 \Gamma_{11} (p) - 2 T \sum\limits_q \Gamma_{11} (q) g_{11} (-p+q), 
	\\ 
	\Sigma_{12} (p) = & n_0 \Gamma_{11} (0), 
\end{align}
where $\Gamma_{11}$ is the $(1,1)$-element of the four-point vertex $\Gamma$ that is connected to the well-known ladder diagrams above $T_{\rm c}$. 

The many-body effect is important for considering the critical temperature shift from that of an ideal Bose gas. 
The critical temperature of the Bose gas in a harmonic trap is known to decrease by the repulsive interaction~\cite{Giorgini1996}. 
This fact can be imagined based on the local density approximation. 
In this approximation, the critical temperature is a monotonic increase function of the peak-density at the center of a harmonic trap. 
In the presence of the repulsive interaction, this density becomes lower, because the repulsive interaction tends to exclude the other bosons. 
This scenario cannot applied to a uniform Bose system, in the case where the system volume and the number of particles are fixed. 
The critical temperature in the uniform system may be shifted by a repulsive interaction in the competition of two effects; the depletion from BEC and the suppression of the fluctuation~\cite{Kashrnikov2001}. 
The depletion may make the critical temperature lower, because the particles are excluded from the condensate by the repulsive interaction. 
The suppression of the fluctuation may make the critical temperature higher. In the first place, an ideal Bose gas has a strong density fluctuation, and its compressibility diverges. 
On the other hand, by the repulsive interaction, the density fluctuation is suppressed, and the compressibility in the repulsively interacting condensed Bose system is finite with satisfying the zero-frequency compressibility sum-rule. 
In the weakly interacting regime, effect of the suppression of the fluctuation is dominant, and the critical temperature increases by the weak repulsive interaction~\cite{Gruter1997, Reppy2000}. 
In the strongly interacting regime, effect of the depletion is dominant, and the critical temperature decreases by the strong repulsive interaction~\cite{Gruter1997, Reppy2000}.

The critical temperature shift in the weak interaction regime is often discussed by the following formula, given by~\cite{Baym2001}
\begin{align}
	\frac{ \Delta T_{\rm c} }{ T_{\rm c}^0 } = - \frac{2}{3} \frac{ \Delta n_{\rm c} (T_{\rm c}^0 )  }{ n_{\rm c}^0 }, 
\end{align} 
where $\Delta T_{\rm c} \equiv T_{\rm c} - T_{\rm c}^0$, and $\Delta n_{\rm c} (T_{\rm c}^0)$ is the difference between the critical density of an interacting Bose gas and that of an ideal Bose gas both at the critical temperature of an ideal Bose gas $T_{\rm c}^0$.  
This difference is given by 
\begin{align}
	& 
	\Delta n_{\rm c}^0 (T_{\rm c}^0) 
		 \label{eq69}
	\\ 
	& 
	\equiv 
	- T_{\rm c} \sum\limits_q 
	\biggl [ \frac{1}{i\omega_n - \epsilon_{\bf q} + \Sigma_{11} (0) - \Sigma_{11} (q)} 
	- 
	 \frac{1}{i\omega_n - \epsilon_{\bf q} } 
	 \biggr ] . 
\nonumber 
\end{align}
Here, we used the Hugenholtz--Pines relation $\mu = \Sigma_{11} (0)$ at the critical temperature, and also used the fact that the chemical potential is zeros at $T_{\rm c}^0$ in an ideal Bose gas. 
If we take the mean-field approximation, such as the Hartree--Fock--Bogoliubov--Popov approximation (Shohno-model), the self-energy $\Sigma_{11}$ does not include the frequency and momentum-dependence. 
As a result, the right hand side of (\ref{eq69}) vanishes, and the mean-field type approximation cannot describe the critical temperature shift~\cite{Baym2001}. 
The critical temperatures in those approximations are the same as that of an ideal Bose gas. 
In order to study the critical temperature shift by the interaction, the frequency or momentum dependent self-energy is needed. 
The critical temperature shift in the dilute limit is characterized by the following equation~\cite{Baym2001,Andersen2004} 
\begin{align}
	\frac{\Delta T_{\rm c}}{T_{\rm c}^0} = 
	c_1 an^{1/3}. 
\end{align}
The values $c_1$ are summarized in table I. 
The Monte-Carlo simulation gives $c_1 \simeq 1.3$~\cite{Arnold2001,Kashrnikov2001}. 
The closet value in our many-body approximation is $c_1 \simeq 1.1$, which is given by the RPA~\cite{Watabe2013}. 
Many other values of $c_1$ are summarized in the review~\cite{Andersen2004}. 

\begin{table}[htb]
 	\begin{tabular}{cccccc}
  	\hline \hline
  			& Monte-Carlo~\cite{Arnold2001,Kashrnikov2001}  & RPA~\cite{Watabe2013} & s-RPA~\cite{Watabe2013} & MBT ~\cite{Watabe2013}
  			     \\   \hline  
  $c_1$ & $1.3$ & $1.1$ & $2.1$ & $3.9$ 
\\       \hline \hline
  \end{tabular}
  \caption{Critical temperature shift $\Delta T_{\rm c} = c_1 a n^{1/3}$, evaluated by numerical calculations: Monte-Carlo simulation~\cite{Arnold2001,Kashrnikov2001}, random-phase approximation (RPA), simplified RPA (s-RPA), and many-body $T$-matrix theory (MBT)~\cite{Watabe2013}. 
  }
  \label{table1}
\end{table}

The density response function can be constructed by the RPA formalism, where the 1PI and 1PR parts are respectively given by~\cite{Watabe2018}
\begin{align}
	\chi^{\rm 1PI} (p) = & \frac{\chi_{\rm R} (p)}{ 1-U\chi_{\rm R} (p) } , 
	\label{eq75}
	\\ 
	\chi^{\rm 1PR} (p) = & {\bf \Upsilon}^\dag (p) {\bf G}(p) {\bf \Upsilon} (p), 
\end{align}
where $\chi_{\rm R}$ is given in (\ref{eq61}), and the density vertices are given by~\cite{Watabe2018} 
\begin{align}
	{\bf \Upsilon} (p) = & \sqrt{-1} [G_{1/2} + {\mathcal G}^\dag \hat T {\bf \gamma} (p) ] A(p) , 
	\\ 
	{\bf \Upsilon}^\dag (p) = & \sqrt{-1} [ G_{1/2}^\dag + {\bf \gamma}^\dag (p) \hat T {\mathcal G} ] A(p) , 
\end{align}
with three point vertices ${\bf \gamma} (p) = \Gamma (p) \Pi (p) | f_0 \rangle$ and ${\bf \gamma}^\dag (p) = \langle f_0 | \Pi (p) \Gamma (p)$. 
Here, we introduced the condensate Green's function $G_{1/2} \equiv \sqrt{-n_0} (1,1)^{\rm T}$, $G_{1/2}^\dag \equiv \sqrt{-n_0} (1,1)$, 
${\mathcal G}_{1/2} = \sqrt{-n_0} \eta_{\rm g}$, and ${\mathcal G}_{1/2}^\dag = \sqrt{-n_0} \eta_{\rm g}^\dag$. 
The matrix $\hat T$ and $\eta_{\rm g}$ are respectively given by~\cite{Watabe2018} 
\begin{align}
	\hat T 
	= 
	\begin{pmatrix}
		1 & 0 & 0 & 0 
		\\ 
		0 & 0 & 1 & 0 
		\\ 
		0 & 1 & 0 & 0 
		\\ 
		0 & 0 & 0 & 1
	\end{pmatrix}, 
	\quad 
	\eta_{\rm g}
	= 
	\begin{pmatrix}
		1 & 0 
		\\
		1 & 0 
		\\ 
		0 & 1 
		\\ 
		0 & 1 
	\end{pmatrix}, 
\end{align}
and $\eta_{\rm g}^\dag$ is the transpose of $\eta_{\rm g}$. 
The vertex coefficient $A (p)$ may be given by~\cite{Watabe2018} 
\begin{align}
	A (p) = \frac{1}{1 - U \chi_{\rm R} (p)}. 
	\label{eq80}
\end{align}
This is one of the way to include infinite series of the selected diagrams into the density response function in the RPA manner for BEC. 
In this case, we find that this density vertex converges a non-zero finite value ${\bf \Upsilon} (0) \neq 0$, which does not satisfy the zero-frequency density vertex identity ${\bf \Upsilon} (0) = 0$. 
In order to satisfy the identity ${\bf \Upsilon} (0) = 0$, we employ the simple replacement of the vertex coefficient $A(p)$ in (\ref{eq80}) by~\cite{Watabe2018} 
\begin{align}
	A(p) = \frac{1}{1 - U \chi_{\rm R}^0 (p)}. 
\end{align}
Since the one-loop approximation $\chi_{\rm R}^0$ shows the infrared divergence, $A (p)$ vanishes in the low-energy limit, which provides the zero-frequency density vertex identity ${\bf \Upsilon} (0) = 0$~\cite{Watabe2018}. 

If the 1PI part is constructed by the infrared divergent one-loop approximation, i.e., $\chi^{\rm 1PI} (p) = \chi_{\rm R}^0 (p) / [1 - U \chi_{\rm R}^0 (p)]$, the 1PI part in the low-energy limit is given by $\chi^{\rm 1PI} (0) = -1/U$. Using the result of the sum-rule $\chi^{\rm 1PI} (0) = - 1/U = - n/(mc^2)$, the sound speed is given by $c = \sqrt{Un/m}$, which is unphysical temperature-independent sound speed. 
The approximation (\ref{eq75}) rather well reproduces the temperature-dependent sound speed $c = \sqrt{-n/[m \chi^{\rm 1PI} (0)]}$, where~\cite{Watabe2018} 
\begin{align}
	\chi^{\rm 1PI} (0) = - \frac{1 - U \Pi'(0)}{3 - 2 U \Pi' (0)}. 
\end{align}

The self-energy in the simplified-RPA reproduces the Nepomnyashchii--Nepomnyashchii identity~\cite{Watabe2013}. 
However, the critical temperature shift $c_1$ in this approximation is over-estimated as shown in table I. 
It will be useful to control the self-energy contributions of the Green's function so as to satisfy  the Nepomnyashchii--Nepomnyashchii identity as well as to reproduce the reasonable value of the critical temperature shift. 
One of the idea is to construct the Green's function similar to that of the Popov's hydrodynamic theory developed in the case at $T = 0$~\cite{Popov1979,Watabe2014}. 

In general, the interaction effect can be included to the Green's function by the Dyson--Beliaev equation (\ref{eq27}), with the irreducible self-energy ${\bf \Sigma}$. 
Another way is to use the non-interacting Green's function ${\bf G}_0$ with the reducible self-energy ${\bf \Sigma}'$, given by 
	\begin{align}
		{\bf G} (p) = {\bf G}_0 (p) + {\bf G}_0 (p) {\bf \Sigma}' (p) {\bf G}_0 (p) , 
	\end{align}
	where 
	\begin{align}
		{\bf \Sigma}' (p) = \frac{1}{1 - {\bf \Sigma} (p) {\bf G}_0 (p) } {\bf \Sigma} (p). 
	\end{align}
The hybrid version~\cite{Watabe2014} 
	\begin{align}
		{\bf G} (p) = \tilde {\bf G} (p) + \tilde {\bf G}_{\rm } (p) \tilde {\bf \Sigma} (p) \tilde {\bf G}_{\rm } (p), 
	\end{align}
	is also possible, 	where $\tilde {\bf G}^{-1} = {\bf G}_0^{-1} - {\bf \Sigma}_{\rm a}$ and $\tilde {\bf \Sigma} = (1-{\bf \Sigma}_{\rm b} \tilde {\bf G} )^{-1} {\bf \Sigma}_{\rm b}$ with ${\bf \Sigma} = {\bf \Sigma}_{\rm a} + {\bf \Sigma}_{\rm b}$. 

The approximation theory satisfying the Nepomnyashchii--Nepomnyashchii identity is constructed by separating the self-energy into two-parts; 
one is the regular part $\tilde {\bf \Sigma}^{\rm R}$, where the bare infrared divergent contribution is not included, and the other is the infrared divergent part $\tilde {\bf \Sigma}^{\rm IR}$~\cite{Watabe2014} . 
We include the regular part into the Green's function by the Dyson--Beliaev type equation, i.e., ${\bf \Sigma}_{\rm a} = \tilde {\bf \Sigma}^{\rm R}$. 
For example, the regular part of the self-energy in the MBT approximation may be given by~\cite{Watabe2014}  
\begin{align}
	\Sigma_{11}^{\rm R} (p) = & 2 n_0 \Gamma_{11}^{R} (p) - 2 T \sum\limits_q \Gamma_{11} (q) g_{11} (-p+q)
	\\ 
	\Sigma_{12}^{\rm R} (p) = & n_0 \Gamma_{11}^{\rm R} (0), 
\end{align}
where $\Gamma_{11}^{\rm R}$ is the $(1,1)$-element of the four-point vertex 
\begin{align}
	\Gamma^{\rm R} (p) = \frac{U}{1 - U \Pi^{\rm R} (p)}, 
\end{align}
where $\Pi^{\rm R} = \Pi - \Pi^{\rm IR}$. 
On the other hand, the infrared divergent part ${\bf \Sigma}^{\rm IR}$ is included into $\tilde {\bf \Sigma}$ within the first order, i.e.,~\cite{Watabe2014}  
\begin{align}
	\tilde {\bf \Sigma} (p) = {\bf \Sigma}_{\rm b} (p) = \tilde {\bf \Sigma}^{\rm IR} (p) . 
\end{align}
In particular, we take the infrared divergent part as~\cite{Watabe2014}  
\begin{align}
	{\bf \Sigma}^{\rm IR} (p) = - \frac{1}{2} G_{1/2} U \langle f_0 | \Pi^{\rm IR} (p) | f_0 \rangle U G_{1/2}^\dag. 
\end{align}
The Green's function in this prescription reproduces the weak infrared divergent longitudinal susceptibility~\cite{Watabe2014}  
\begin{align}
	\chi_{\parallel} (p) \simeq & - \frac{n_0 U^2 \Pi_{14} (p)}{2 [\Sigma_{12}^{\rm R} (p)]^2 }. 
\end{align}
Given the approximated Green's function in the form ${\bf G} = \tilde {\bf G} + \tilde {\bf G}\tilde {\bf \Sigma}\tilde {\bf G}$, we obtain the self-energy ${\bf \Sigma}$ by solving the Dyson--Beliaev equation (\ref{eq27}), given in the form~\cite{Watabe2014}  
\begin{align}
	{\bf \Sigma} = & {\bf G}_0^{-1} 
	- 
	\frac{1}{1 + \tilde {\bf \Sigma} \tilde {\bf G}} \tilde {\bf G}^{-1}. 
\end{align}
Since $\tilde {\bf \Sigma} = {\bf \Sigma}^{\rm IR}$ shows the infrared divergence, 
we have ${\bf \Sigma} (0) = {\bf G}_0^{-1} (0) = \mu$. 
It indicates that the present prescription can reproduce the Hugenholtz--Pines relation, as well as the Nepomnyashchii--Nepomnyashchii identity. This formulation is consistent with the Popov's hydrodynamic theory at $T=0$, where the Green's function is given by the correlation functions of the hydrodynamic variables. 

\section{Summary}

In this paper, starting from the discussion of properties of an ideal Bose gas, exact relations in an interacting condensed Bose gas is discussed. 
In particular, we focused on the Hugenholtz--Pines relation, Nepomnyashchii--Nepomnyashchii identity, and the density response function. 
The Hugenholtz--Pines relation is related to the gapless excitation of the single-particle excitation. 
The Nepomnyashchii--Nepomnyashchii identity gives a weak infrared divergence of the longitudinal susceptibility. 
The zero-frequency compressibility sum-rule is exhausted by the one-particle irreducible part of the density response function. 
On the other hand, the one-particle reducible part of the density response function, which is specific to the BEC, has the same pole of the single-particle Green's function, whose sound speed is equal to that of the thermodynamic sound mode.  

In the last part of this paper, we discussed the approaches of many-body approximations satisfying these exact relations. We introduced two random phase approximations and the many-body $T$-matrix theory. 
We discussed the critical temperature shift by the many-body effect in these approximations, and the systematic approach to reproduce the Hugenholtz--Pines relation, Nepomnyashchii--Nepomnyashchii identity, as well as the weak-infrared divergence of the longitudinal susceptibility. 
We also discussed the random phase approximation for studying the density response function with satisfying the identity. 
There is not a many-body approximation satisfying all the exact relations in a Bose--Einstein condensate (BEC). 
This paper will be useful for developing beyond mean-field theory of an interacting condensed Bose system consistent with exact relations in BECs. 


%
%
%
%
%
%
%
%

%


\acknowledgments
The author thanks Y. Kato, Y. Ohashi, M. Ueda, and T. Nikuni for valuable discussions and comments on BEC and superfluidity. 
A series of the author's study on BECs are supported by JSPS KAKENHI Grant No. (249416, JP16K17774, JP18K03499). 
He also gratefully acknowledges the support by Nanotech Career-up Alliance (Nanotech CUPAL), Japan Science and Technology Agency (JST).




\end{document}